# PVLAS Results

Ugo Gastaldi, INFN-LNL Legnaro
Viale dell'Università 2, 35020 Legnaro (Pd) Italy
representing the PVLAS Collaboration[1]

**Abstract**
The PVLAS experiment has built an apparatus to measure rotations and ellipticities induced by a transverse magnetic field onto linearly polarized laser light. Available results can be interpreted as observation of an unexpectedly large rotation effect between $10^{-7}$ and $4 \; 10^{-7}$ rad. If this rotation is due to the existence of a light spin-zero boson of mass m and coupling to two photons 1/M, the ball-park values of the boson mass parameters are m = 1 meV and M = $10^6$ GeV.

---

[1] PVLAS Collaboration (Udine, Trieste, Pisa, Legnaro, Frascati, Ferrara): E. Zavattini, G. Zavattini, G. Ruoso, E. Polacco, G. Petrucci, E. Milotti, M. Karuza, U. Gastaldi, G. Di Domenico, F. DellaValle, R. Cimino, S. Carusotto, G. Cantatore, M. Bregant.

# PVLAS Results

Ugo Gastaldi, INFN-LNL Legnaro
Viale dell'Università 2, 35020 Legnaro (Pd) Italy
representing the PVLAS Collaboration[2]

**Abstract**
The PVLAS experiment has built an apparatus to measure rotations and ellipticities induced by a transverse magnetic field onto linearly polarized laser light. Available results can be interpreted as observation of an unexpectedly large rotation effect between $10^{-7}$ and $4\,10^{-7}$ rad . If this rotation is due to the existence of a light spin-zero boson of mass m and coupling to two photons 1/M , the ball-park values of the boson mass parameters are m = 1 meV and M = $10^6$ GeV.

**1-Introduction**

The PVLAS experiment has been built to probe quantum vacuum with linearly polarized laser light[1]. Physics informations are to be extracted from observations and measurements of induced ellipticity and/or apparent rotation of the polarization plane of light emerging from the apparatus after that an initially linearly polarized light beam has traversed a vacuum region with an intense transverse magnetic field.

For these purposes a 1.3 m long superconducting dipole magnet provides a 1 m long 5.5 T uniform magnetic field, and a tunable CW Nd:YAG laser provides infrared laser light (λ= 1064 nm, ω~1.2 eV). A Fabry-Perot cavity (FP) embracing the magnet is used to increase the number N of passages of the photons of the laser beam through the magnetized region. A heterodyne technique is used to enhance the signal over background ratio.

The original motivation[2] of the experiment was to observe directly at a macroscopic level the ellipticity induced by vacuum polarization from electron-positron loops (fig 1a). Incoming linearly polarized light is expected from QED to emerge from the apparatus with an induced ellipticity $\Psi_{QED}$ ~ $3.6\,10^{-16}$ rad per passage trough the magnet.

Photon splitting is a QED source of rotation of the polarization plane (fig.1b). In a very intense magnetic field the incoming polarized photon of energy ω may split into a pair of real collinear photons $\omega_1$ and $\omega_2$ with $\omega = \omega_1 + \omega_2$. This process depends on the photon polarization and can generate rotation by differential absorption of the two polarization states parallel and normal to

---

the magnetic field. It is expected to be extremely rare[3] with magnetic fields **B** obtainable in a laboratory, since it is proportional to a factor $(B/B_{crit})^6$, where $B_{crit} = 4.4 \cdot 10^{13}$ Gauss.

Another source of induced ellipticity would be the existence of spin-zero light particles with a coupling to two photons that would permit virtual transitions as indicated in fig. 1d to occur[4]. If spin-zero light particles exist and their mass m is lighter than the laser photon energy ω, also real transitions could occur (Primakoff effect, fig.1c), and photon spin-zero-boson oscillations would be possible inside the magnet[4,5]. This oscillation would cause depletion of the component of the photon beam with polarization vector **E** parallel to the magnetic field **B** for pseudoscalars (and with **E** orthogonal to **B** for scalars), and generate therefore an apparent rotation ρ of the polarization plane. The ellipticity Ψ and rotation ρ due to the existence of a light spin-zero particle depend on its mass m and its coupling to two photons $g_{\gamma\gamma}=1/M$. The signs of rotation for scalar and pseudoscalar spin-zero particles are opposite[4].

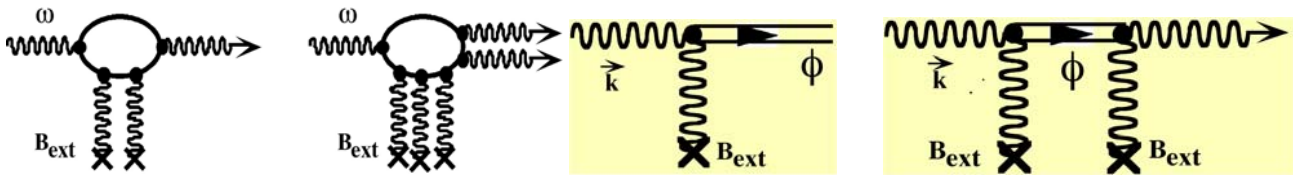

Fig.1 QED vacuum polarization graph (a), photon splitting (b), spin-zero boson real production (c), spin-zero boson virtual production (d).

One could also speculate about photon absorption by two photon coupling to a spin 2 field. However the coupling is likely to be so faint to render negligible the process in comparison to the others above mentioned.

PVLAS has started data taking in 2001 and has reached a sensitivity of about $10^{-7}$ rad Hz$^{-1/2}$ with N ~ $10^5$ traversals of the 1 m long 5.5 T magnetized region. A $10^3$ sec long run enables therefore observation of a $10^{-7}$ rad ellipticity or rotation effect as a 10σ effect, if a $10^{-11}$ rad ellipticity or rotation is induced per passage through the magnet. The noise level has usually been around $10^{-8}$ rad for a $10^3$ sec long run.

In runs lasting typically $10^3$ sec we have always observed in ellipticity and rotation measurements effects at the level from several $10^{-8}$ to some $10^{-7}$ rad. These observations are rather clear for rotation measurements performed during 2004, that we are in the process to submit for publication[6], while for ellipticity further analysis of the data appears necessary.

The amplitude of the observed signals is rather surprising. It exceeds by 4 orders of magnitude QED expectations for induced ellipticity and, if the observed signals of induced rotations are interpreted as due to a light spin zero particle, it suggest values for the parameters m and M respectively around $10^{-3}$eV and $10^{6}$GeV. The value of m is in the window open for axion[7] searches, while 1/M is about 6 orders of magnitude away from expectations of the coupling to two photons for a $10^{-3}$eV axionlike particle[8,9]. The observed signals point to new physics, if they are not generated in the apparatus by normal processes that we have not spotted.

We mention very briefly the experimental method and the apparatus, that are covered in previous publications[1], and concentrate on calibrations and rotation measurements. In the outlook we mention plans of measurements to assess the results and establish dependences of the observed effects on physical parameters like the laser photon energy $\omega$ and the magnet length L, which control the reactions and can be changed in planned steps.

**2- Apparatus and experimental method**

The apparatus extends vertically. The dipole magnet is installed on a rotating table and has a cylindrical bore aligned with the table axis. The magnetic field lines in the bore rotate on horizontal planes. The rotation of the magnetic field modulates (for the purpose of the heterodyne technique) the ellipticity and apparent rotation effects under study at twice the rotation frequency $\omega_m$ of the turntable. The interferential mirrors of the Fabry-Perot cavity are 6.4 m distant from each other and are installed in vacuum boxes, that are well above and below the magnet and are connected by a quartz tube which traverses the magnet bore. The bottom vacuum chamber contains a polarizer P before the FP mirror. In the top vacuum chamber, after the FP mirror, there is a quarter wave plate (QWP) followed by an ellipticity modulator and by an analizer A oriented normally to P. A diode D detects the light transmitted by the analyzer A. The ellipticity modulator is a Stress Optical Modulator[10] SOM used as carrier in the heterodyne technique. The SOM is driven by a precise sine generator at 506 Hz, and introduces an ellipticity of amplitude $2.5 \cdot 10^{-3}$ that beats with the effect of the magnetic field modulated at twice the turntable rotation frequency. The investigated signals appear as $2\omega_m$ sidebands of the carrier SOM modulator frequency $\omega_S$. The signal of the photodiode D is sampled at a frequency of 8.2 kHz by a large dynamical range ADC. A frequency analysis of the signals of D features peaks at multiples m of the carrier modulator frequency $\omega_S$ and their $n\omega_m$ sidebands. The ellipticity and rotation effects under study are derived from the amplitudes and phases of the peaks in the Fourier analysis of the signals of the detection diode. Optics and magnet have no mechanical interconnections and are supported by two separate foundations 14 m deep in the ground, in order to minimize couplings that can feed the $2\omega_m$ sidebands of the $\omega_S$ peak.

**3- Calibrations**

Linearly polarized light that traverses a medium in presence of a transverse magnetic field acquires an elliptical polarization. This effect, known as Cotton-Mouton effect (CME)[11] is very small in gases and its amplitude is proportional to the gas pressure at fixed temperature. The ellipticity acquired by an initially linearly polarized light has maximum amplitude when the magnetic field direction is at $45^0$ to the initial polarization. PVLAS can measure directly amplitude and phase of the ellipticity due to CME when gas is present in the optical cavity simply by removing the QWP. We have measured the dependence of the CME on the intensity of the magnetic field B by performing sequential runs under the same conditions of the optics and powering the magnet with several current values. These measurements integrate also the effects of the magnetic fringe field at the two ends of the magnet. We have measured CME in $N_2$ and Ne, Ar, Kr, Xe[12]. All measurements show good proportionality between gas pressure and CME signal amplitude. Measurements with Ne have shown that the B dependence is very well described by a $B^2$ curve. Fig. 2 shows in a polar plot signal amplitude (as radius) and phase of measurements with $N_2$ and Ne at various pressures: one may notice that the phase is independent on the gas pressure, but that, as known from the literature, Ne and $N_2$ feature opposite sign CME, and therefore opposite phases. The QWP does not completely suppress ellipticity generated in the FP cavity. The residual ellipticity generated by CME of gas in the optical cavity can be used to calibrate directly in amplitude and phase the ellipsometer when it is in the configuration with QWP inserted into the light beam path. We have checked that after a rotation by $90^0$ of the QWP the phase of the residual ellipticity signal is shifted by $180^0$. By comparing the residual rotation signal as a function of Ne pressure and the ellipticity signal in runs of December 2004 we observe that the slope of the rotation curve is about 10 times smaller than the slope of the ellipticity curve, that both ellipticity and rotation data point to an intercept with the zero pressure line that is different from zero and that the phases of both rotation and ellipticity signals change when the pressure is reduced below 3 mbar. This indicates that at least another source of rotations and ellipticities becomes noticeable below 3mbar.

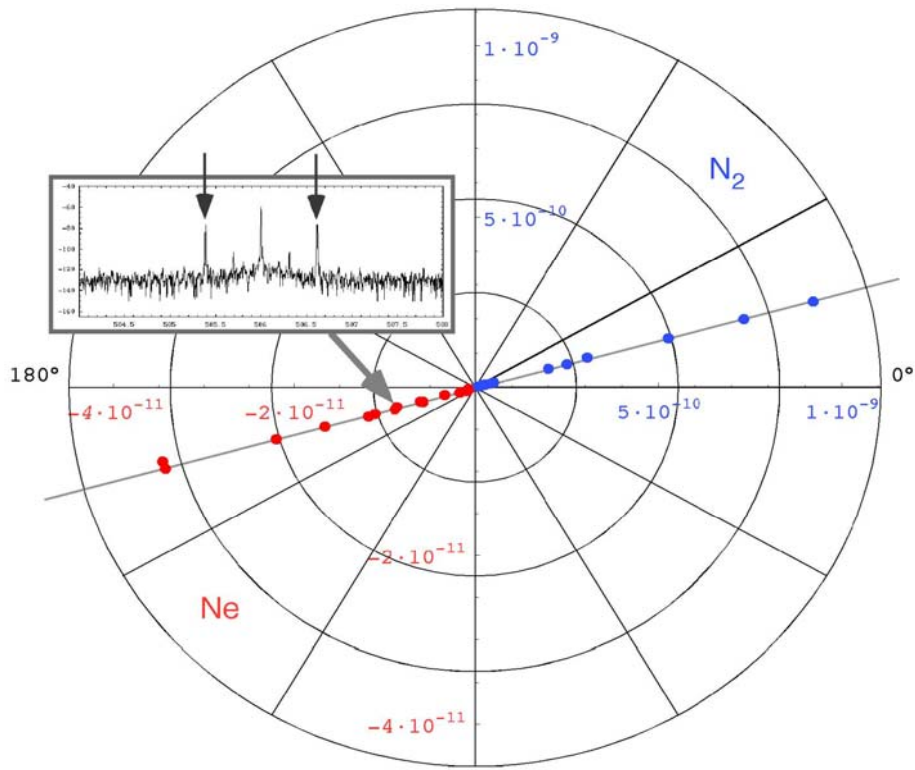

Fig. 2 Polar plot of signal amplitude (as radius) and phase for CME measurements with $N_2$ and Ne at several pressure values. Each dot represents a run at one pressure value.

**4- Rotations of the polarization plane in vacuum**

In the Fourier amplitude spectra of runs with vacuum in the quartz tube and nominal magnetic field B = 5.5T the $\omega_m$ sidebands of $\omega_s$ are the dominant feature of the spectrum and the $2\omega_m$ sidebands of $\omega_s$ are well visible. This situation is typical of all the runs performed with vacuum in the interaction point and 5.5T magnetic field. With residual gas pressure below $10^{-7}$mbar, no visible signal is expected at $\omega_s \pm 2\omega_m$ from CME of the residual gas. The $\omega_m$ and $2\omega_m$ sidebands of $\omega_s$ are instead absent when the magnet is off.

In order to identify the source of the signals at $\omega_s \pm 2\omega_m$, we have performed several runs changing the QWP from a $0^0$ position to a $90^0$ position in order to check whether the signal is due to pure optical rotation. We have changed the intensity of the magnetic field in a series of runs performed without acting on the optical regulations, in order to establish the signal dependence on the intensity of the magnetic field, and have verified that the dependence is well described by a $B^2$ curve (in agreement with expectations for a rotation signal generated in the magnet) for sets of sequential runs. We have also checked, by removing one interferential mirror of the cavity, that the signal originates from the cavity and it is not due to some kind of electronic pick-up. It has furthermore been checked that the signal intensity is proportional to the finesse of the FP.

The $\omega_m$ sideband is present at B=5.5T in measurements with gas in the quartz tube, but its amplitude does not depend on the gas pressure. It appears associated to the presence of the magnetic field and to its action on parts of the apparatus outside the interaction region in the magnet. The mechanical effects of rotation of the magnet with field off do not generate the $\omega_m$ nor the $2\omega_m$ sidebands, since these sidebands are absent with B=0 and magnet in rotation.

Fig. 3 shows data points of two polar plots (amplitude, phase) from 2 pairs of runs (2 runs per plot) in vacuum taken sequentially with B=5.5.T and with QWP set at $0^0$ and $90^0$ respectively. These data were collected in 2004 after completion of an improvement program of the vacuum system. In these plots each point represent the amplitude and phase of the $2\omega_m$ sidebands of $\omega_s$ of a 100 sec long subset of a data acquisition run. The phase expected for optical rotations (and confirmed by calibration runs with gases shown in fig.2) is at $15^0$ at the centre of the plots from the horizontal line. These data show clearly that the $2\omega_m$ sidebands are largely due to an optical rotation effect, since the $90^0$ and $0^0$ clusters of points are at $180^0$ to each other in all runs, as expected from the action on an optical ellipticity or rotation signal from the rotation by $90^0$ of the QWP.

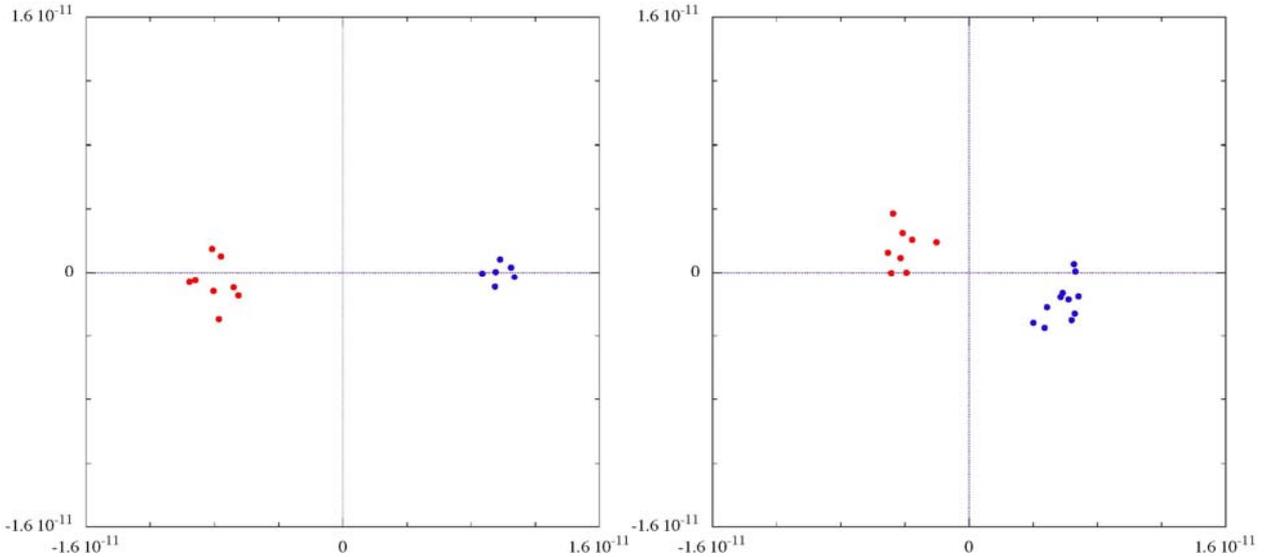

Fig.3 Amplitude and phase points representing signals of the $2\omega_m$ sideband of $\omega_s$ for 100 sec long subsets of data acquisition runs with QWP at $90^0$ (left points) and QWP at $0^0$ (right points).

During all 2004 the points of pairs of sequential runs with $90^0$ and $0^0$ QWP are always well separated in phase and amplitude, the lines that ideally join the centres of gravity of the points of a pair of clusters are not completely aligned with the optical axis (identified by the calibration runs with gases), and the amplitude of the centre of gravity of each cluster varies from about $2\ 10^{-12}$ to $8\ 10^{-12}$ rad per light passage.

These features indicate that at least two effects contribute to generate the $2\omega_m$ lateral bands of $\omega_s$. One of them acts along the optical axis and one has a different phase. The vector combination of these effects (all modulated at $2\omega_m$) produces a global result that appears stable during one run, while a variation of the amplitude of one (at least) of the effects should cause the instability around the optical axis, when moving from one pair of runs to another pair performed in a different period of time or after interventions on the optics.

We are investigating correlations between signals at $2\omega_m$ and presence of $3\omega_m$ and $4\omega_m$ sidebands of $\omega_s$ and presence of $\omega_m$ and $2\omega_m$ sidebands of $2\omega_s$, and also correlations with the phase behaviour of the $\omega_m$ sideband of $\omega_s$, in order to improve data selection.

At present the global features of the data are compatible with a scenario where the $2\omega_m$ sideband of $\omega_s$ is generated by optical rotations in vacuum plus at least another effect on the optics generated directly by the residual magnetic field onto materials of the optics away from the magnet, and/or by mechanical movements induced on optics elements by the forces associated to the rotating stray magnetic field, and/or by modulated movements or intensity of the laser beam.

The optical rotation due to vacuum effects has its phase along the optical axis. Its amplitude can therefore be obtained by projecting onto the optical axis half the vector obtained by the difference of the vector signals associated to $90^0$ and $0^0$ QWP runs, under the hypothesis that no other optical rotation acts along the optical axis (this is not the case e.g. if gas is present in the vacuum enclosure). The ball park value derived this way for vacuum rotations $\rho$ is of the order of $10^{-7}$ to $4\ 10^{-7}$ rad.

**5- Conclusions**

We observe a clear signal in the $2\omega_m$ sidebands of the $\omega_s$ peak in vacuum measurements, which features $B^2$ dependence on the magnetic field and has a dominant component along the optical axis of the (amplitude, phase) plane. If this is not an artefact of the apparatus, we are observing a rotation of the polarisation plane of the light injected into the apparatus due to selective absorption of the component of light with polarization parallel to the magnetic field direction. In other words, this corresponds to observation of magnetically induced dichroism of vacuum. A preliminary analysis of our data with various Ne gas pressures also support the hypothesis that the signal observed in vacuum is mainly due to magnetically induced dichroism of vacuum[13].

The most plausible reason for the disappearence of the (extremely little) fraction of photons with polarization component parallel to the magnetic field is the production of pseudoscalar particles with mass m smaller than the energy $\omega$ of the laser photons by photon-photon interaction of laser photons with virtual photons of the magnetic field.

The rotation effect $\rho$ caused by the existence of a pseudoscalar particle of mass m and two photon coupling 1/M is given in vacuum by

$$\rho = N\ B^2\omega^2 M^{-2} m^{-4} \sin^2(m^2 L/4\ \omega)$$

The existence of a pseudoscalar would also be responsible for an induced ellipticity given by

$$\Psi = \tfrac{1}{2}\ N\ B^2\omega^2 M^{-2} m^{-4}\ [m^2 L/2\ \omega - \sin(m^2 L/2\ \omega)]$$

The value for induced rotations is in the window $10^{-7} - 4\ 10^{-7}$ rad. The ball park value for induced ellipticity in PVLAS is in the window $10^{-8} - 2\ 10^{-7}$ rad.

The above windows define two strips in the (m,M) plane that intercept each other in an area around the m = 1meV, M = $10^6$GeV point. This area is at the edge of the region excluded by the former laser experiment of the BFST Collaboration[14] and quite in the middle of a large domain excluded by astrophysical and cosmological arguments. While m = 1meV is in the region where existence of axions is still expected, the 1/M coupling constant is 6 orders of magnitude larger than expected in most recent axion models.

The values of induced ellipticity and dichroism in vacuum depend both significantly and in different ways on the energy ω of the laser photons and on the magnet length L. We plan to run in the near future measurements with laser energy doubled, to see if the rotation effect in vacuum is confirmed and to check if the values of rotation and ellipticity obtained at two laser energies are consistent with the hypothesis of existence of a light pseudoscalar particle.

Later on we will remove the access structure to PVLAS, that is made of iron tubes and install in place a new one made in aluminium. We contemplate on a longer time scale to make measurements with a set of modular permanent magnets that would permit to measure rotations and ellipticities with 4 different values of the magnet length.

If the rotation signal and the pseudoscalar hypothesis will be confirmed by all those future measurements, the present values of m and M are such that a regeneration experiment[15] using LHC type magnets could be performed with the purpose of assessing definitively the production and observation in the lab of a new type of particles likely to represent components of dark matter.